\documentstyle[aps,prl,twocolumn,epsfig]{revtex}

\begin{document}
\draft

\twocolumn[\hsize\textwidth\columnwidth\hsize\csname@twocolumnfalse\endcsname
\title{Charge Ordering and Spin Dynamics in NaV$_{2}$O$_{5}$}
\author{B. Grenier$^{a}$, O. Cepas$^{b}$, L.P. Regnault$^{a}$, J.E. Lorenzo$^{c}$,
T. Ziman$^{b}$, J.P. Boucher$^{d}$, A. Hiess$^{b}$,\\ T.
Chatterji$^{b}$, J. Jegoudez$^{e}$ and A. Revcolevschi$^{e}$}
\address{$^{a}$D\'{e}partement de Recherche Fondamentale sur la Mati\`{e}re
Condens\'{e}e, SPSMS, Laboratoire de Magn\'{e}tisme et de
Diffraction Neutronique, CEA-Grenoble, F-38054 Grenoble cedex 9,
France
\\$^{b}$Institut Laue Langevin, BP 156, F-38042 Grenoble cedex 9,
France.\\ $^{c}$Laboratoire de Cristallographie, CNRS, BP 166,
F-38042 Grenoble cedex 9, France.\\ $^{d}$Laboratoire de
Spectrom\'{e}trie Physique, Universit\'{e} J. Fourier Grenoble I,
BP 87, F-38402 Saint Martin d'H\`{e}res cedex, France.\\
$^{e}$Laboratoire de Physico-Chimie de l'Etat Solide,
Universit\'{e} Paris-Sud, B$\hat{a}$t 414, F-91405 Orsay cedex,
France\\}
\date{ June 2000}
\maketitle
\begin{abstract}
We report high-resolution neutron inelastic scattering experiments
on the spin excitations of NaV$_{2}$O$_{5}$. Below $T_{c}$, two
branches with distinct energy gaps are identified. From the
dispersion and intensity of the spin excitation modes, we deduce
the precise {\it zig-zag} charge distribution on the ladder rungs
and the corresponding charge order: $\Delta _{c}\approx 0.6$. We
argue that the spin gaps observed in the low-T phase of this
compound are primarily due to the charge transfer.

\end{abstract}

\pacs{PACS numbers: 71.45.Lr, 75.10.Jm, 75.40.Gb}
\
] \narrowtext The low dimensional inorganic compound
NaV$_{2}$O$_{5}$ undergoes a phase transition at $T_{c}=34$
K\cite{Isobe} associated with both a lattice
distortion\cite{Chatterji} and the opening of an energy gap to the
lowest triplet spin excitations\cite{Yosihama}. While the nature
of the low-$T$ phase in NaV$_{2}$O$_{5}$ is not fully understood
it is clear that, unlike CuGeO$_{3}$, the spin-Peierls model does
not apply simply to this compound \cite{Bompadre}. The spin gap
may result from charge-order (CO) rather than the lattice
distortion\cite {Mostovoy}. Indeed, NMR measurements indicate two
inequivalent vanadium sites below $T_{c}$, while there exists only
one site above\cite {Ohama 99}. There has been no direct evidence
for the connection between CO and a spin gap, nor to distinguish
various conjectured spatial distributions of charge
\cite{Mostovoy,Thalmeier/Yaresko}. In this letter, we present new
results of neutron inelastic scattering (NIS) on the spin
excitations in the low-$T$ phase that can now resolve these
issues.

In NaV$_{2}$O$_{5}$, the vanadium ions have a formal valence of
4.5+. Initially, this was proposed to correspond to an alternation
of V$^{4+}$ ions, with a spin value $S=1/2$, and V$^{5+}$ ions
with $S=0$\cite{Carpy}. At room temperature, NaV$_{2}$O$_{5}$ is
well described by a quarter-filled two-leg ladder system, with
only one type of vanadium site V$^{4.5+}$. From calculations of
electronic structure\cite{Smolinski,Suaud}, the strongest orbital
overlaps are on the ladder rungs. One expects that the $S=1/2$
spins are carried by the V-O-V molecular bonding orbitals, with
charge fully delocalized on two sites. As the energy of the
anti-bonding orbital is much higher, it can be projected out, and
above $T_{c}$, these spins, as they interact in the leg direction
(${\Vert }$ ${\bf b}$ axis), form an effective uniform quantum
Heisenberg spin {\it chain} with interactions between chains that
are both weaker and frustrated.

At low temperatures, NMR shows this can no longer be so. On each
rung, a charge transfer $\Delta _{c}$ may occur. Taking the
average charge on vanadium sites to be 1/2, the charges on the two
vanadium sites on a rung are defined through $n_{\pm }=(1\pm
\Delta _{c})/2$. Two forms of CO can be considered\cite{Mostovoy},
the {\it in-line}, with the same charge transfer on each rung, and
the {\it zig-zag} with alternation in the charge along the ladders
as shown in fig. 1b. Recent X-ray diffraction
measurements\cite{Ludecke} established that the lattice structure below $%
T_{c}$ consists of a succession of distorted and non-distorted
ladders of
vanadium ions (see fig. 1a). Neglecting inter-ladder diagonal couplings $%
J_{\bot }$, the ladders would behave magnetically as independent
spin chains. For one ladder (chain 2 in fig. 1a) distortions in
the exchange paths both within the ladder and via neighboring
ladders result in an alternation of the effective exchange
coupling in the {b} direction, $J_{b1}$ and $J_{b2}$. The ladders
in which the rungs are distorted (chains 1 and 3), however, remain
magnetically uniform as a mirror plane passes through each rung. A
{\it minimum} magnetic model without CO would be a succession of
alternating and uniform chains. An energy gap (expected to be
small as it results primarily from the alternation in $J_{\bot }$)
would characterize the excitation branch of the alternating
chains, and there would be no gap for the uniform chains. The
initial NIS\cite{Yosihama} found, however, two
excitation branches, with the same gap at the antiferromagnetic point $%
E_{g}^{+}=E_{g}^{-}\approx 10$ meV. To analyse these results, a
recent spin model\cite{Gros} used an explicit relation between the
spin excitations and the CO. This model assumed a single gap and,
as it implies zero intensity of one excitation branch, must be
extended to explain the NIS data. A more precise determination of
excitations in the low-T phase of NaV$_{2}$O$_{5}$ is therefore
crucial. In the present work, using high-resolution, the
dispersion of the excitations is re-explored in a wider part of
the reciprocal space. Moreover, the evaluation of the structure
factor, i.e., the (energy-integrated) intensity of each excitation
mode, allows us to determine the charge transfer $\Delta _{c}$.

The single crystal ($\approx 8$x$5$x$2$ mm$^{3}$) was grown by a
flux method. The NIS measurements were performed at $T\leq 4.2$ K,
on two thermal neutron three-axes spectrometers - IN8 and
CRG/CEA-IN22 - at the Institut Laue-Langevin (ILL). On IN8,
vertically focusing monochromator PG($002$) and Cu($111$) were
used in conjunction with a vertically focusing analyzer PG($002$)
and horizontal collimations $60^{\prime }$-$40^{\prime
}$-$60^{\prime } $. The final wave vector was kept fixed at
$k_{f}=4.1$ \AA $^{-1}$. IN22 was operated at $k_{f}=2.662$ \AA
$^{-1}$, with a PG($002$) monochromator and a PG($002$) analyzer
used in horizontal monochromatic focusing condition (resulting
wave vector resolution: $\delta q\approx 0.03$ r.l.u.) with no
collimation. The sample was installed in an ``orange'' ILL
cryostat, with the scattering wave-vector ${\bf Q}$ lying in the
reciprocal ($a^{*},b^{*}$) ladder plane.

\begin{figure}[tbp]
\centerline{\epsfxsize=70mm \epsfbox{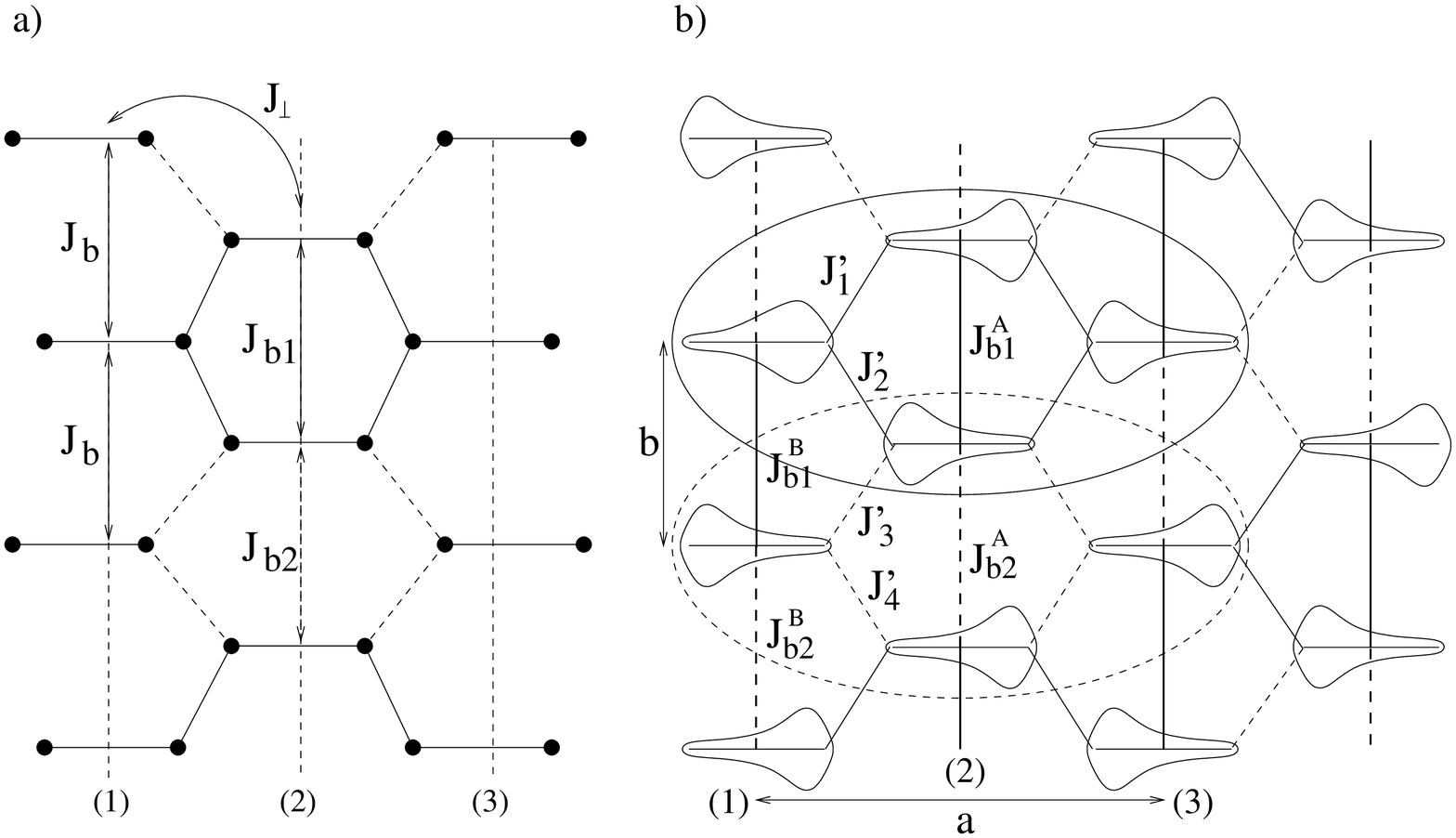}} \vspace{+0.2
cm} \caption{a) Simplified representation of distorted and
non-distorted chains, chain 1 (or 3) and 2, respectively, with the
$J_{\bot }$ bond alternation between chains 1 and 2; b) Proposed
charge order (large and small lobes represent large and small
average charges) leading to a spin gap. The elementary translation
(${\bf a}$,${\bf b}$) of this CO agrees with the observed
periodicity of spin excitations.}
\end{figure}

The two branches characterizing the low-energy excitations in NaV$_{2}$O$%
_{5} $ have distinct energy gaps: $E_{g}^{+}\neq E_{g}^{-}$. This
important result will be established when we consider the
dispersions in the transverse {\bf a} direction. First, however,
we determine the dispersion in the leg direction ({\bf b} axis).
Examples of constant-energy scans obtained on IN8 as a function of
$Q_{b}$ are displayed in fig. 2a (wave-vector components are
expressed in reciprocal lattice units, r.l.u.). Increasing the
energy, one resolves the single peak seen at 10 meV into
propagating modes (dashed lines), whose peak position is shown in
fig. 2b. They describe the dispersion of the elementary
excitations in the {\bf b} direction, near the AF chain
wave-vector component $Q_{b}^{AF}$ ($\equiv Q_{b}=0.5$). They
are compared to a dispersion law characteristic of a gapped spin chain: $%
E(q_{b})=\sqrt{E_{g}^{2}+(E_{m}^{2}-E_{g}^{2})\sin ^{2}(2\pi Q_{b})}$ where $%
E_{g}$ and $E_{m}$ are the gap and maximum energies of the
dispersion, respectively. For both $E_{g}=E_{g}^{+}$, $E_{g}^{-}$
(the solid and dashed lines, respectively), we evaluate
$E_{m}=93\pm 6$ meV. Compared to the prediction for a uniform
Heisenberg chain, $E_{m}=\pi J_{b}/2$, where $J_{b}
$ is the exchange in the chain, one obtains for the low-$T$ phase, $%
J_{b}\approx 60$ meV.

Second, we consider the dispersions in the {\bf a}* direction
(i.e., along the rungs). A few examples of energy scans performed
on IN22 at constant ${\bf Q}$ are reported in fig. 3. In general,
two peaks are observed. At a few $Q_{a}$ values, however, an
extinction occurs. This extinction may concern one of the two
modes, or the two modes
simultaneously. The left and right panels report data obtained for $%
Q_{b}^{AF}=0.5$ and $Q_{b}=1$ (equivalent to the zone center of
the AF chains $Q_{b}^{ZC}$). As examples, we show for $Q_{b}^{AF}$
that the smallest energy difference between the two observed peaks
is obtained for
{\it integer} $Q_{a}$ values (here, $Q_{a}=3)$, the largest one for {\it %
half-integer} values ($Q_{a}=2.5)$ and an extinction of the two
modes occurs
at $Q_{a}\approx 1.75$ (identical results have been obtained for $Q_{b}=1.5$%
). Surprisingly, at the chain zone-center $Q_{b}^{ZC}$ we found a
small but non-zero intensity for the two excitation branches. The
smallest energy difference between the two peaks is now obtained
for {\it half-integer} values (here, $Q_{a}=1.5$). Extinctions of
one of the two modes is observed at $Q_{a}=1$ (on the high-energy
mode) and at $Q_{a}=2$ (on the low-energy mode). For all the
spectra recorded on IN22, the background (the dotted lines in fig.
3) was carefully determined. Several procedures have been used
involving data recorded at low and high temperature (i.e., above
$T_{c}$). In fig. 3a, for instance, the open dots used to define
the background are obtained from measurements performed at 40
K\cite{Background} while in fig. 1b, it is determined from $Q_{b}$
scans performed at low temperature for different energies.

\begin{figure}[tbp]
\centerline{\epsfxsize=74mm \epsfbox{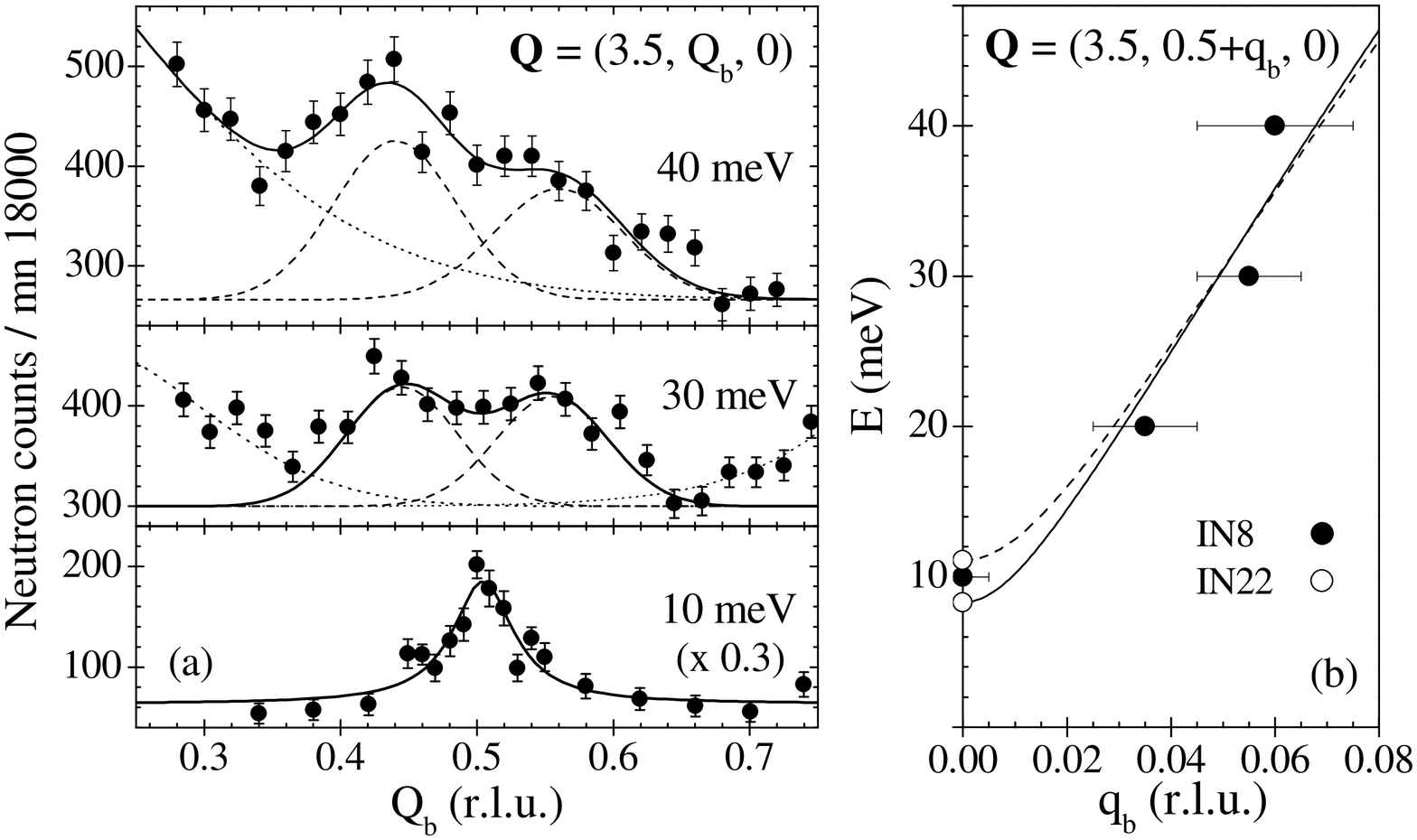}} \vspace{-0.8
cm} \caption{IN8 data: a) Constant-energy scans as a function of
$Q_{b}$. The full lines is the result of a fit to the data, which
takes into account both the magnetic contributions (shown as the
dashed lines) and the background (the dotted line); b) Energy
dispersion in the ${\bf b}$ direction. The lines correspond to the
two branches with gaps $E_{g}^{+}$ and $E_{g}^{-}$.}
\end{figure}

For the analysis, we assume the two observed peaks to belong to
two distinct contributions. Their unsymmetrical lineshape is
characteristic of gapped excitations undergoing a rapid energy
dispersion (as established in fig. 2b). In such a case, a
dynamical response function (shown by the dashed lines in fig. 3)
is well-suited to fitting, conveniently defined with only 3
parameters: the peak energy $E_{+}$ ($E_{-}$), an intensity factor $A_{+}$ ($%
A_{-}$) and an energy damping $\Gamma $\cite{Mason}. As $\Gamma $
is mainly fixed by the resolution conditions, it is assumed to be
the same for the two contributions ($\Gamma \approx 0.4-0.8$ meV).
Together with the background, the agreement with the experiments,
shown by the solid lines, is good. The values obtained for $E_{\pm
}$ as a function of $Q_{a}$ are shown in fig. 4.
We establish several new features. At both $Q_{b}^{AF}$ (solid symbols) and $%
Q_{b}^{ZC}$ (open symbols), the transverse dispersion consists of
two distinct excitation branches which never cross. This justifies
our previous statement, namely that there are {\it two} distinct
energy gaps, $E_{g}^{+}$ and $E_{g}^{-}$. In each branch, the
periodicity is $2\pi Q_{a}$: this is twice that previously
determined. The two dispersions have the same amplitude, $\delta
J\approx 1$ meV but, remarkably, the upper and lower
branches are out of phase, and there is phase inversion between branches at $%
Q_{b}^{ZC}$ and $Q_{b}^{AF}$. For each excitation branch, the
corresponding structure factors, $S_{b\pm }^{AF}(Q_{a})$ and
$S_{b\pm }^{ZC}(Q_{a})$ are evaluated by integrating the fitted
dynamical response functions over a wide energy range (from $0$ up
to $E\approx 300\Gamma )$. We estimate systematic error for
varying the upper cut off alters the results by at most 5\%. The
resulting values (dots and squares) are reported in figs. 5a and
b. The sums
$S_{b}^{AF}(Q_{a})=S_{b+}^{AF}(Q_{a})+S_{b-}^{AF}(Q_{a})$ and $%
S_{b}^{ZC}(Q_{a})=S_{b+}^{ZC}(Q_{a})+S_{b-}^{ZC}(Q_{a})$ are shown
as the stars.

\begin{figure}[tbp]
\centerline{\epsfxsize=73mm \epsfbox{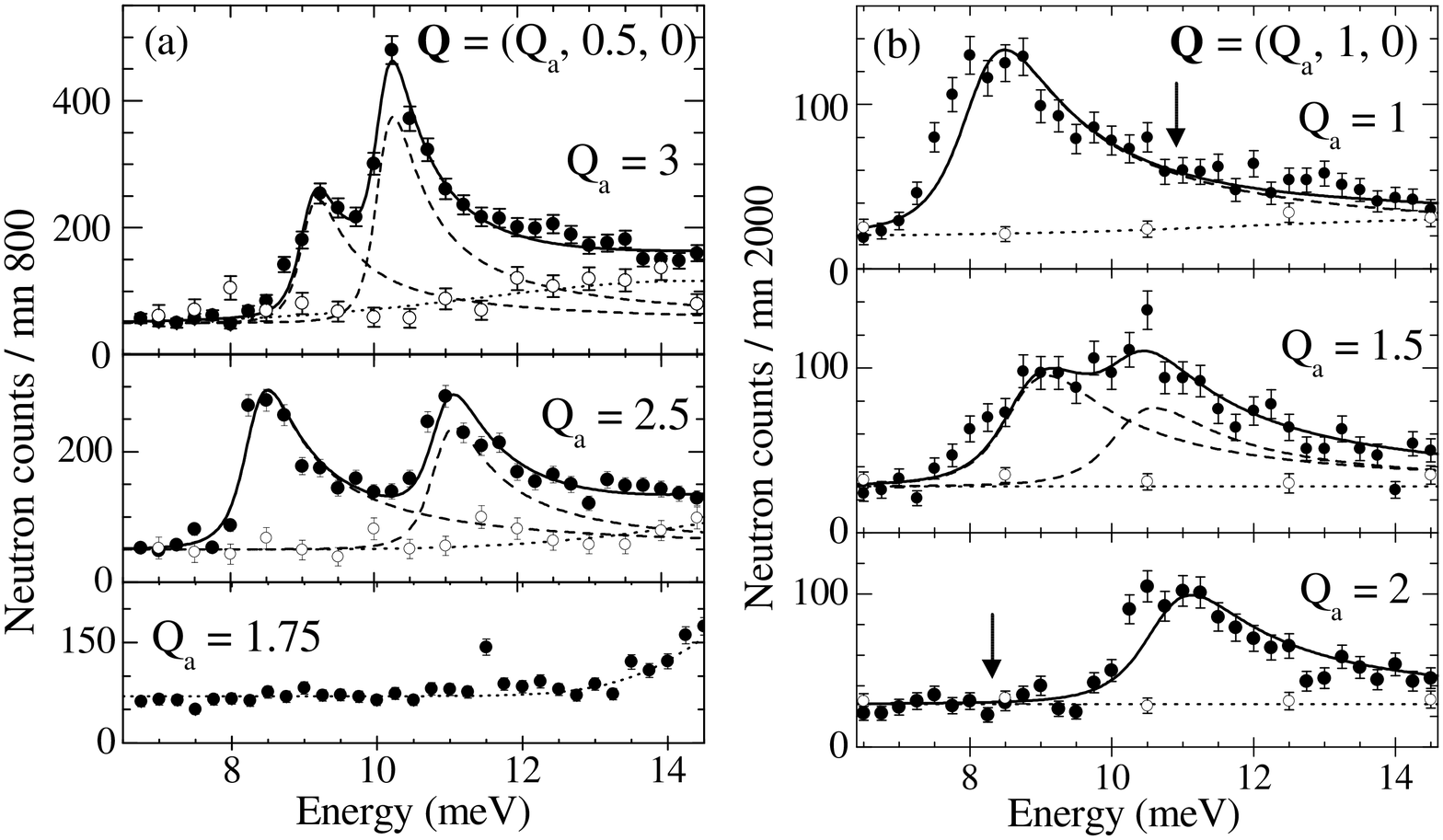}} \vspace{-0.7
cm}
\caption{IN12 data: Constant-${\bf Q}$ scans as function of energy. a) at $%
Q_{b} \equiv Q_{b}^{AF}$; b) at $Q_{b} \equiv Q_{b}^{ZC}$. The two
arrows point to the positions expected for the two missing peaks.
Curves are described in text.}
\end{figure}

The interpretation of these results is developed in three steps.
First, the charge order: each spin is associated with an
electronic wave function on the two sites of a rung that depends
on $n_{\pm }$. The structure factors for the {\it in-line} and
{\it zig-zag} models are $S_{Q_{b}}(Q_{a},\omega )=\cos ^{2}\left(
\pi Q_{a}\rho \right) \tilde{S}\left( Q_{a},Q_{b},\omega
\right) +\Delta _{c}^{2}\sin ^{2}\left( \pi Q_{a}\rho \right) \tilde{S}%
\left( Q_{a}+{\frac{1}{2}},Q_{b},\omega \right) $ and $S_{Q_{b}}(Q_{a},%
\omega )=$ $\cos ^{2}\left( \pi Q_{a}\rho \right) \tilde{S}\left(
Q_{a},Q_{b},\omega \right) $ + $\Delta _{c}^{2}\sin ^{2}\left( \pi
Q_{a}\rho \right) \tilde{S}$($Q_{a}+{\frac{1}{2}}$,
$Q_{b}+{\frac{1}{2}},\omega $),
respectively, with $\omega =E/\hbar $ and where $\rho =l/a\approx 0.304$ ($l$%
, rung length and $a$, lattice parameter) and $\tilde{S}\left(
Q_{a},Q_{b},\omega \right) $ is the structure factor for spins
localized on the center of each rung\cite{Cepas}. From the ratios
of intensities for different values of momentum, one can extract
the charge transfer $\Delta _{c}$ independent of the form of
$\tilde{S}$. In particular, we verify that the order cannot be
{\it in-line} but predictions agree with a {\it zig-zag} order
with $\Delta _{c}^{2}\approx 0.35$, i.e., $\Delta _{c}\approx
0.6$.
The agreement is particularly good for the sums $S_{b}^{AF}(Q_{a})$ and $%
S_{b}^{ZC}(Q_{a})$. For the absolute intensities, we need an
explicit form for $\tilde{S}$ which we take from the strongly
dimerized limit (SDL), in which the wave function is simply a
product of singlets on the stronger bonds. For the {\it in-line}
model, for example, the SDL would give zero intensity at
$Q_{b}^{ZC}$ in contradiction with the observation (the data in
fig. 5b). The {\it in-line} model can be ruled out. In figs. 5a
and b, the predictions provided by the {\it zig-zag} model (solid
lines) are compared
with the experimental total structure factors $S_{b}^{AF}(Q_{a})$ and $%
S_{b}^{ZC}(Q_{a})$ (solid and open stars). In fig. 5a, the
agreement is obtained with no adjustable parameter except for an
overall amplitude factor. Once this factor is determined, the
results in fig. 5b depends only
on $\Delta _{c}^{2}$. As can be seen, a good agreement is obtained for $%
\Delta _{c}^{2}\approx 0.35$. The low-T phase of NaV$_{2}$O$_{5}$
is very well described by the zig-zag model, with a rather large
charge transfer.

\begin{figure}[tbp]
\centerline{\epsfxsize=73mm \epsfbox{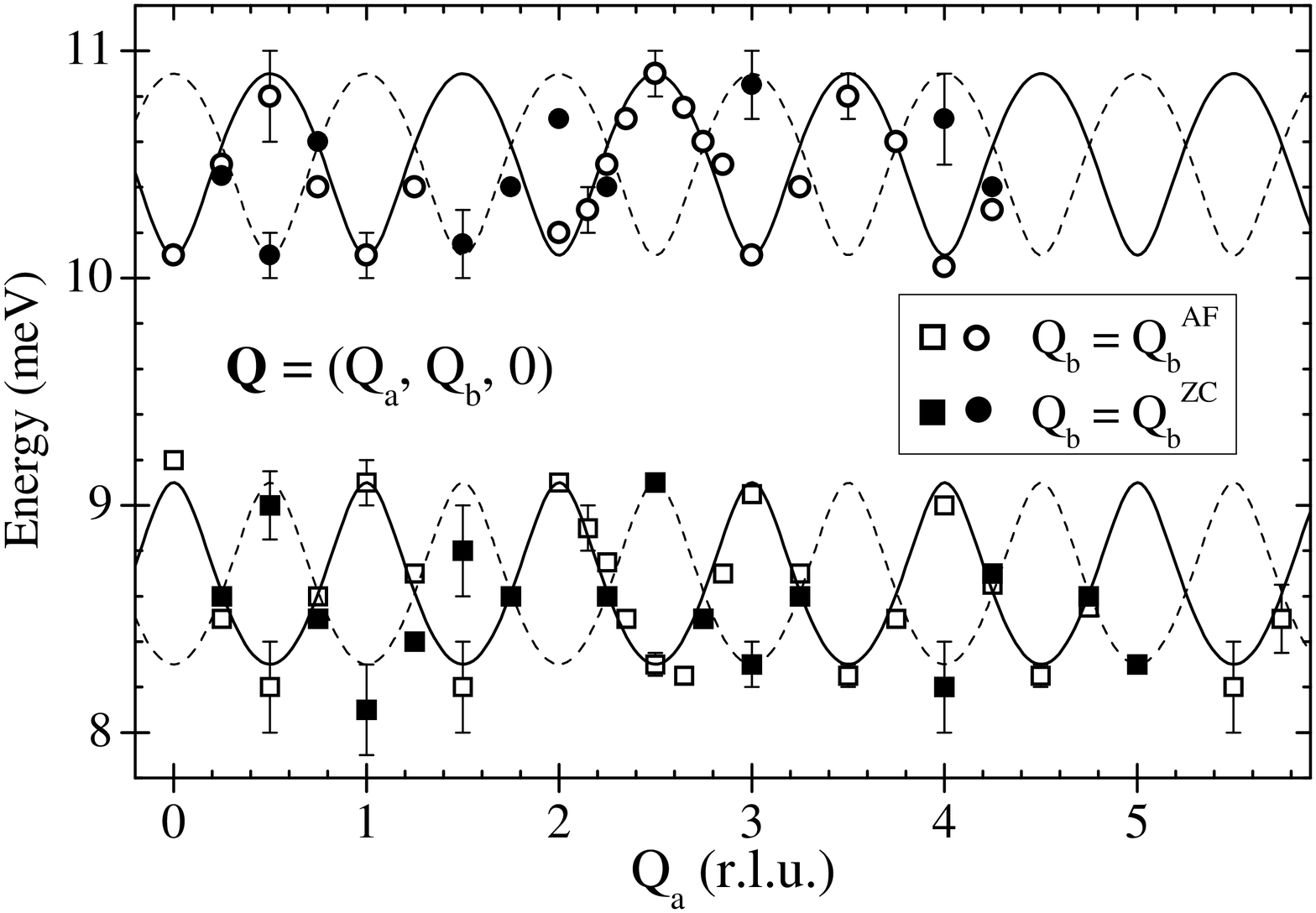}} \vspace{+0.3
cm} \caption{Dispersion of the two excitation branches in the
${\bf a}$ direction at $Q_{b}^{AF}$ and $Q_{b}^{ZC}$. The curves
are theoretical predictions fitted to the data (see text). }
\end{figure}
Second, we consider the transverse dispersions. Later, we explain
that the gap is induced by the CO. In fact, there are two distinct
gaps because of
the structural distortions (implying distinct couplings $J_{b1,2}^{A}$ and $%
J_{b1,2}^{B}$ as shown in fig. 1b). Due to the CO, 4 different
interchain exchange integrals must be considered, $J_{1}^{\prime
}$, $J_{2}^{\prime }$, $J_{3}^{\prime }$ and $J_{4}^{\prime }$.
The two branches (associated with the gaps $E_{g}^{+}$ and
$E_{g}^{-}$) acquire a transverse dispersion, described at
$Q_{b}^{AF}$ and $Q_{b}^{ZC}$ by $E_{b\pm }^{AF}=\left(
E_{g}^{+}+E_{g}^{-}\right)\!\!/2\pm \sqrt{\left(
E_{g}^{+}-E_{g}^{-}\right) ^{2}\!\!/4+\delta J^{2}\sin ^{2}\left(
\pi Q_{a}\right) }$ and $E_{b\pm }^{ZC}=\left(
E_{g}^{+}+E_{g}^{-}\right) \!\!/2\pm \sqrt{\left(
E_{g}^{+}-E_{g}^{-}\right) ^{2}\!\!/4+\delta J^{2}\cos ^{2}\left(
\pi Q_{a}\right) }$, respectively, with $\delta J=J_{1}^{\prime
}-J_{2}^{\prime }+J_{3}^{\prime }-J_{4}^{\prime }$\cite{Cepas}. In
fig. 4, these predictions (solid and dashed lines) are compared to
the data. Again, a very good agreement is obtained yielding the
following evaluation $E_{g}^{+}=10.1\pm 0.1$, $E_{g}^{-}=9.1\pm
0.1$ and $\delta J=1.2\pm 0.1$ meV. The dispersion
gives directly the {\it alternation} in the inter-ladder diagonal bonds $%
\delta J$. If we assume it is dominated by the CO, we can also
estimate the {\it average} from $\delta J\approx J_{\bot }\Delta
_{c}^{2}$\cite{Gros} giving $J_{\bot }\approx 2.4$ meV.

The structure factors can be calculated for each branch.\ Within
the SDL approach, one obtains the contributions $S_{b\pm
}^{AF}(Q_{a})$ and $S_{b\pm }^{ZC}(Q_{a})$\cite{Struct.} shown by
the dotted and dashed lines in fig. 5. In fig. 5b, the agreement
with the data is rather good. In particular, the extinction
phenomenon observed for each branch is well reproduced. In fig.
5a, while the sum $S_{b}^{AF}(Q_{a})$ is well described, we note a
discrepancy between the SDL predictions and the {\it individual}
structure factors $S_{b+}^{AF}(Q_{a})$ and $S_{b-}^{AF}(Q_{a})$.
This difficulty could be explained as follows. As shown in fig.
1b, two successive chains are not identical. A charge transfer
giving different average valence on the chains
will mix intensities at $Q_{b}^{AF}$ without affecting the fluctuations at $%
Q_{b}^{ZC}$ in qualitative agreement with the
observation\cite{Remark.}. Experimental supports for such a charge
transfer would be useful. Our proposition for the charge ordering,
i.e., the {\it zig-zag} model sketched in fig. 1a, and our
estimate $\Delta _{c}\approx 0.6$ are based on the {\it total}
intensities.

\begin{figure}[tbp]
\centerline{\epsfxsize=80mm \epsfbox{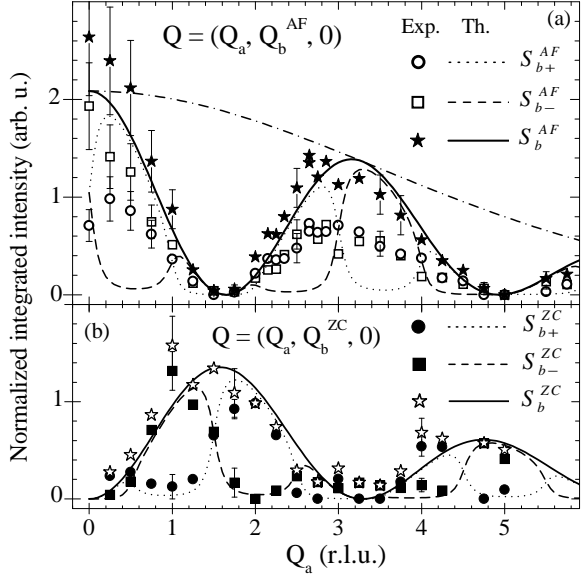}} \vspace{- 3.2
cm} \caption{Structure factors $S_{b\pm }^{AF}(Q_{a})$ and
$S_{b\pm }^{ZC}(Q_{a}) $ for the two magnetic branches in the
${\bf a}$ direction: a) at $Q_{b}^{AF} $; b) at $Q_{b}^{ZC}$. The
solid and open stars represent the sums $S_{b}^{AF}(Q_{a})$ and
$S_{b}^{ZC}(Q_{a})$. The data are corrected
from the V$^{4+}$ atomic form factor $f_{V^{4+}}$, being all normalized at $%
Q_{b}=0.5$. The dot-dashed line gives the $Q_{a}$ dependence of $f_{V^{4+}}$%
. The other curves are theoretical predictions compared to the
experiments (see text).}
\end{figure}

Finally, we consider the origin of the gap. As discussed in the
introduction, the lattice distortion alone, in isolated ladders,
cannot explain the presence of two energy gaps and their size. To
analyze the effects of the diagonal couplings $J_{\bot }$, we
refer to fig. 1b. Each ladder is seen to be a succession of two
distinct clusters (shown by ovals in the figure). By exact
diagonalization of each cluster, using the effective parameters of
a t-J model \cite{Suaud} and adding a potential imposing a charge
transfer, we evaluated $J_{b1}$ and $J_{b2}$ as a function of
$\Delta _{c}$ and the bond alternation $d=\left|
J_{b1}-J_{b2}\right| /\left( J_{b1}+J_{b2}\right) $. For $\Delta
_{c}\approx 0.6$, the couplings underestimate the experimental
value by a factor of about 2, but as the parameters calculated on
the high temperature structure \cite{Suaud} and as such cluster
calculations are rather crude, the agreement is satisfactory. The
value obtained for the bond alternation can be considered as
reasonable: $d\approx 0.025-0.030$. Then, using the experimental
value $J_{b}\approx 60$ meV, one finds an energy gap $E_{g}\approx
6-8$ meV. This is in a fairly good agreement with the experimental
value $E_{g}\approx 10$ meV. This simple analysis supports the
view that, in NaV$_{2}$O$_{5}$, the gaps are primarily due to the
CO\cite{Ohama 00}. The lattice distortion plays a secondary role,
explaining why two distinct branches are observed experimentally,
and their separation. In our picture, the magnetic anisotropies
are unnecessary\cite {Thalmeier/Yaresko}.

In conclusion, the CO in NaV$_{2}$O$_{5}$ is quantitatively
determined by the present NIS measurements \cite{Trebst}. It
explains also the energy gaps observed in the low-T phase of this
compound.


\begin{references}

\bibitem{Isobe}  M. Isobe and Y. Ueda, J. Phys. Soc. Jpn {\bf 65}, 1178
(1996).

\bibitem{Chatterji}  T. Chatterji et al., Solid State Com. {\bf 108}, 23
(1998).

\bibitem{Yosihama}  T. Yosihama et al., J. Phys. Soc. Jpn {\bf 67}, 744
(1998).

\bibitem{Bompadre}  S.G. Bompadre et al., cond-mat/9911298.

\bibitem{Mostovoy}  Mostovoy and Khomski, cond-mat/9806215; Damascelli et
al. Phys. Rev. Lett. {\bf 81}, 918 (1998). P. Thalmeier and P. Fulde, Europhys. Lett., {\bf 44}%
, 142 (1998); H. Seo and K. Fukuyama, J. Phys. Soc. Jpn {\bf 67},
2602 (1998); J. Riera et al., Phys. Rev. B {\bf 59}, 2667 (1999);
A.I. Smirnov et al., {\it ibid} {\bf 59}, 14546 (1999); Schwenk et
al., {\it ibid} {\bf 60}, 9194 (1999); M. Lhomann et al., Phys.
Rev. Lett. {\bf 85}, 1742 (2000).

\bibitem{Ohama 99}  T. Ohama et al., Phys. Rev. B {\bf 59}, 3299 (1999).

\bibitem{Thalmeier/Yaresko}  P. Thalmeier and A.N. Yaresko, Eur. J. Phys. B
{\bf 14}, 495 (2000).

\bibitem{Carpy}  P.A. Carpy and J. Galy, Acta Crystallogr B {\bf 31}, 1481
(1975).

\bibitem{Smolinski}  H. Smolinski et al., Phys. Rev. Lett. {\bf 80}, 5164
(1998).

\bibitem{Suaud}  N. Suaud and M.B. Lepetit, Phys. Rev. B {\bf 62}, 402
(2000).

\bibitem{Ludecke}  J. L\"{u}decke et al., Phys. Rev. Lett. {\bf 82}, 3633
(1999).

\bibitem{Gros}  C. Gros and R. Valenti, Phys. Rev. Lett. {\bf 82}, 976
(1999).

\bibitem{Background} Taking into account an estimated magnetic
contribution for this temperature.

\bibitem{Mason}  T.E. Mason et al., Phys. Rev. Lett. {\bf 69}, 490
(1992).

\bibitem{Cepas}  O. Cepas, PhD. Thesis, Grenoble, France,2000.

\bibitem{Struct.}  The corresponding expressions agrees with the sum rules $%
S_{b}^{AF}(Q_{a})$ and $S_{b}^{ZC}(Q_{a})$ established above.

\bibitem{Remark.}  In contrast, if we assume different values of $\Delta
_{c}$ on two successive chains, the individual modes will mix at
$Q_{b}^{ZC}$ rather than at $Q_{b}^{AF}$ unlike what is observed.

\bibitem{Ohama 00}  In T. Ohama et al., cond-mat/0003141, the CO proposed in
model Z2 (fig. 3) does not give rise to energy gaps, model Z3
provides a gap for only one out of two chains, model Z1 is
magnetically equivalent to ours (fig. 1b).

\bibitem{Trebst} Other models assuming three different valence states (V$^{4+}$,
V$^{4.5+}$ and V$^{5+}$) have been shown to be incompatible with
our results: S. Trebst et al, Phys. Rev. B {\bf 62}, R14613
(2000); C. Gros et al., {\it ibid} {\bf 62}, R14617 (2000); and
references therein.

\end{references}
\end{document}